\documentstyle[12pt]{article}
\begin{document}
\begin{flushright}
October 1991 \\
EFI 91-58\\
\end{flushright}

\bigskip
\begin{center}
TOPOLOGICAL FIELD THEORIES AND THE SPACE-TIME SINGULARITY \\

\bigskip
\medskip

{\it Tohru Eguchi } \\
\bigskip

{\it Enrico Fermi Institute, University of Chicago} \\
{\it 5630 S.Ellis, Chicago, Illinois 60637, U.S.A.} \\~\\
{\it and} \\~\\
{\it Department of Physics, Faculty of Science} \\
{\it University of Tokyo, Tokyo, JAPAN 113}

\end{center}

\newcommand {\beq}{\begin{equation}}
\newcommand {\eeq}{\end{equation}}

\centerline{ABSTRACT}

\begin{quote}
Based on a study of recently proposed solution of $2$ dim. black hole
we argue that the space-time singularities of general relativity may be
described by topological filed theories (TFTs). We also argue that in general
TFT is a field theory which describes singular configurations of a reduced
holonomy in its field space.
\end{quote}
\newpage
\vspace{10mm}

	Recent discovery of a 2 dim. black hole being described by an exactly
soluble 2 dim. conformal field theory \cite{W1} offers exciting possibilities:
some fundametal issues of the theory of general relativity such as the
space-time
singularities and Hawking radiation may be studied from a totally new
point of view.

	In this article we would like to address ourselves with the problem
of singularities and raise the possibility that the singularities of general
relativity may be described topological field theories (TFTs).
As Witten has already noted in his paper, the gauged $SL(2;R)$ WZW model is
well-defined at the 2 dim. black hole
singularity and approaches a TFT of a vanishing $U(1)$
gauge connection near the singularity. In the following we shall constrcut
a TFT which is closely related to the $SL(2;R)$ gauged WZW model and
show that it
describes the region only around the singularity. We also argue that
in general when the
target space of TFT has a non-trivial geometry and contains singularities,
the path integral of TFT receives dominant contributions from the
singularities. Thus TFT is a field theory which
describes the structure of singular configurations in its field space.

	Lets us now start with the discussion with the $SL(2;R)$ gauged
WZW model. Its action is given by
\begin{eqnarray}
&&S_{WZW}=\frac{k}{8\pi}\int Tr\partial_{i}g\partial_{i}g^{-1} d^{2}z +
i\frac{k}{12\pi} \int \epsilon^{ijk} Tr (g^{-1}\partial_{i}g
g^{-1}\partial_{j}g g^{-1}\partial_{k}g) d^{3}x  \nonumber \\
&& +\frac{k}{2\pi}\int Tr(A_{\overline{z}}\sigma_{3}g^{-1}\partial_{z}g +
A_{z}\sigma_{3}\partial_{\overline{z}}gg^{-1} + A_{z}\sigma_{3}
A_{\overline{z}}\sigma_{3} +
A_{z}\sigma_{3}gA_{\overline{z}}\sigma_{3}g^{-1})d^{2}z. \nonumber  \\
\label{eq:1}
\end{eqnarray}

Here $g$ is a $2\times2$ matrix parametrized as
\beq
g=\left( \begin{array}{rc}
 a & u \\
 -v & b
\end{array} \right),~~~ab+uv=1
\label{eq:1a}
\eeq
and $\sigma_{3}$ is the Pauli matrix.
The action is invariant under the $U(1)$ gauge transformation
\begin{eqnarray}
& &\delta g=\epsilon(\sigma_{3}g + g\sigma_{3}), ~~~
\delta A_{i}=-\partial_{i}\epsilon,  \nonumber \\
& &(\delta a=2\epsilon a,~~\delta b=-2\epsilon b,~~\delta u=\delta v=0).
\label{eq:2}
\end{eqnarray}

In the region $1-uv>0$ a suitable gauge condition is $a=b$ while in the
region $1-uv<0$, $a=-b$. Written in terms of functions $u,v,a,b$ the above
action is expressed as

\begin{eqnarray}
&&S_{WZW}=-\frac{k}{4\pi}\int (\partial_{z}u\partial_{\overline{z}}v +
\partial_{\overline{z}}u\partial_{z}v + \partial_{z}a\partial_{\overline{z}}b
+ \partial_{\overline{z}}a\partial_{z}b)d^{2}z  \nonumber \\
&&+\frac{k}{2\pi}\int (A_{\overline{z}}(b\partial_{z}a-a\partial_{z}b+
u\partial_{z}v-v\partial_{z}u) + A_{z}(b\partial_{\overline{z}}a -
a\partial_{\overline{z}}b - u\partial_{\overline{z}}v + v\partial_
{\overline{z}}u) \nonumber \\
& &+4A_{z}A_{\overline{z}}(1-uv) +
\ln a (\partial_{z}u\partial_{\overline{z}}v-\partial_{\overline{z}}u
\partial_{z}v))d^{2}z.
\label{eq:3}
\end{eqnarray}

If we use the gauge condition (\ref{eq:2}) and eliminate the gauge potential
$A_{z},
A_{\overline{z}}$ by means of the equation of motion, we obtain (up to
 $1/k$ corrections)
\beq
S=-\frac{k}{4\pi}\int d^{2}x \frac{\partial_{i}u\partial_{i}v}{1-uv}.
\label{eq:4}
\eeq

Thus the model describes a $2$ dim. target space geometry with the metric
\beq
ds^{2}=\frac{dudv}{1-uv}.
\label{eq:6}
\eeq

Eq.(\ref{eq:6}) gives a $2$ dim. black hole. The curvature of the metric blows
up at $uv=1$. Thus the space-time
singularities appear at uv=1 while the event horizons occur at uv=0.
It is, however, easy to recognize that the gauged WZW model itself
is well-defined everywhere including the singularities. Only the process
of gauge fixing and the elimination of the gauge field breakes down
at $uv=1$. If one parametrizes $u$ and $v$ as $u=e^{w},v=e^{-w}$ near the
singularity and rewrite the action, one finds
\beq
S=-\frac{k}{4\pi}\int d^{2}x D_{i}aD_{i}b + i\frac{k}{2\pi}\int d^{2}x
w \epsilon^{ij}F_{ij}.
\label{eq:7}
\eeq

The second term in the above action is in fact a simple TFT of
the vanishing $U(1)$ gauge connection (the 3 dim. $U(1)$ Chern-Simons gauge
theory dimensionally reduced to $2$ dimension.)
 On the other hand the first term
of the action describes the Gaussian fluctuation of the fields $a,b$ in
the transverse direction to the singularity. Thus the gauged WZW model
reduces to some simple TFT near the singularity.

(We note that both $a$ and $b$ vanish at the singularity.
This is easy to see in a parametrization
of $SL(2;R)$ using Euler angles. For instance, in the region between
the (future) singularity and (future) horizon $a,b,c,d$ are parametrized as
\beq
\begin{array}{ll}
 a=cosre^{\sigma},& b=cosre^{-\sigma}                  \\
 u=sinre^{t},& v=sinre^{-t}
\end{array},~~0\leq r\leq\pi/2,~~t,\sigma\in R.
\label{eq:7a}
\eeq
$\sigma$ is a gauge-dependent parameter and is shifted as $\sigma \rightarrow
\sigma +2\epsilon$ under (\ref{eq:2}). Using the gauge freedom we can always
bring $\sigma$ to a finite value. Then $a$ and $b$ vanish at the same time
at the singularity $r=\pi/2$).

At the singularity the matrix $g$ is reduced to a form
\beq
g = \left( \begin{array}{rr}
            0 & u    \\
            -v & 0
           \end{array}  \right),~~ uv=1
\label{eq:8a}
\eeq
and the symmetry currents
of the model $g^{-1}D_{z}g$,$D_{\overline{z}}gg^{-1}$ have
an Abelian structure
\begin{eqnarray}
g^{-1}D_{z}g = \left( \begin{array}{cc}
                       u\partial_{z}v & 0 \\
                       0  &  v\partial_{z}u
                      \end{array} \right),  \label{eq:16a} \\
D_{\overline{z}}gg^{-1} = \left( \begin{array}{cc}
                       v\partial_{\overline{z}}u & 0  \\
                       0  &  u\partial_{\overline{z}}v
                                 \end{array} \right). \label{eq:16b}
\end{eqnarray}
Thus the original $SL(2;R)$ symmetry of the theory is reduced to $U(1)$
symmetry. This subspace in the field space of the reduced symmetry (holonomy)
corresponds
to the black hole singularity in the space-time interpretation. We also
note that the geometry of $2d$ black hole is covered by two gauge patches
$uv<1$,$uv>1$ and the singularity occurs at their border. The singularity
itself is invariant under the transformation (\ref{eq:2}) and thus forms
a fixed point of $U(1)$ gauge transformation.

It is quite curious to see if the association of a TFT to the space-time
singularity is a generic phenomenon and we can provide some physical
arguments to such an association.
In order to study this problem let us first
consider a TFT which is closely related to the gauged $SL(2;R)$ WZW model.
We shall see that our toplogical model describes only
the region of the singularity in the black hole geometry.

Our model is obtained first by making the $SL(2;R)/U(1)$ WZW model
supersymmetric by adding fermions belonging to the coset $SL(2;R)/U(1)$.
Supersymmetric $SL(2;R)/U(1)$ model then has an extended $N=2$ superconformal
symmetry since the coset $SL(2;R)/U(1)$ has a complex structure
\cite{KS}. We can then twist the $N=2$ superconformal symmetry \cite{EY,W2}
and obtain a TFT which
is closely related to the original $SL(2;R)/U(1)$ WZW model.

The action of the superconformal version of the $SL(2;R)/U(1)$ model
is given by
\beq
S^{super}_{WZW}=S_{WZW} + \frac{i}{2\pi}\int d^{2}z
(Tr\Psi D_{\overline{z}} \Psi + Tr\overline{\Psi} D_{z}\overline{\Psi})
\label{eq:8}
\eeq
where $\Psi,\overline{\Psi}$ are fermions belonging to the coset $SL(2;R)/U(1)$
\beq
\Psi(z)=\left( \begin{array}{cc}
  0 & \psi(z) \\
 \psi^{*}(z) & 0
\end{array} \right), ~~~~
\overline{\Psi}(\overline{z})=\left( \begin{array}{cc}
  0 & \overline{\psi}^{*}(\overline{z})  \\
  \overline{\psi}(\overline{z}) & 0
\end{array} \right).
\label{eq:9}
\eeq

It is easy to see that the action (\ref{eq:8}) possesses an extended $N=2$
superconformal symmetry \cite{Nak} in accordance with the general theorem
\cite{KS}.
The process of twisting $N=2$ symmetry in $2$ dimensions amounts to
a mixing of Lorentz and internal $U(1)$ rotations and redefinition of
the quantum numbers of the fields \cite{W2,EY}. After twisting one of the
supersymmetry
operators is turned into the BSRT operator and we obtain a theory with
BRST gauge symmetry. The new stress tensor in the twisted theory has
a vanishing central charge. Physical states (BRST invariants) are given by
the chiral ring \cite{LVW} of $N=2$ theory. In the present case
the fermions are converted into spin $0,1$ ghost fields $\alpha$ and $\beta$
by twisting.
The action of our topological model then is given by
\beq
S^{t}_{WZW}=S_{WZW} + \frac{i}{2\pi}\int d^{2}z Tr (\beta_{z}D_{\overline{z}}
\alpha + \overline{\beta}_{\overline{z}}D_{z}\overline{\alpha}).
\label{eq:10}
\eeq

(\ref{eq:10}) is the type of the model recently studied by Witten
in connection with $2d$ gravity and the $N$ marix model \cite{W3}.
In \cite{W3} it was shown
that in a model with a fermionic gauge symmetry dominant contributions to its
path integral come from the fixed points of the gauge transformation. This is
due to the fact that the volume of the gauge orbit of a fermionic gauge
symmetry
is proportional to a factor
\beq
\int d\theta =0
\label{eq:11}
\eeq
where $\theta$ is an anti-commuting c-number. Thus the contribution vanishes
if the transformation acts freely without fixed points. Therefore the dominant
contribution to the path integral of (\ref{eq:10}) is
expected to come from the fixed points of the BRST transformation. We note
that the BRST fixed points are field configurations which are annihilated
by the BRST transformation and thus are the natural path-integral analogue
of the physical state condition in the operator formulation
\beq
Q|physical\rangle =0
\label{eq:18}
\eeq
where $Q$ is the BRST operator.

Let us now study the BRST symmetry of (\ref{eq:10}). The BRST transformation
directly follows from the supersymmetry transformation of (\ref{eq:8}) and
is given by \cite{W3,Nak}
\begin{eqnarray}
&&\delta g= i\epsilon(\alpha g \sigma_{3} + \overline{\alpha} \sigma_{3} g),
\label{eq:12}  \\
&&\delta \alpha = \delta \overline{\alpha} = \delta A_{i} = 0,
\label{eq:13}   \\
&&\delta \beta_{z} = \frac{1}{2}\epsilon
Tr((\sigma_{1}+i\sigma_{2})g^{-1}D_{z}g),
\label{eq:14}   \\
&&\delta \overline{\beta}_{\overline{z}} = \frac{1}{2}\epsilon
Tr((\sigma_{1}-i\sigma_{2})
D_{\overline{z}}gg^{-1}).
\label{eq:15}
\end{eqnarray}
Since the anti-commuting fields are put to zero when we look for the BRST
fixed points, essentail conditions come from the last two equations. We note
that their right hand sides vanish when the currents $g^{-1}D_{z}g$,
$D_{\overline{z}}gg^{-1}$ have no off-diagonal elements, i.e. the $SL(2;R)$
symmetry is reduced to the Abelian $U(1)$ symmetry.
We note that this is exactly the same condition as we have seen before in
connection with the fixed points of the $U(1)$ gauge transformation (eqs.
\ref{eq:16a}, \ref{eq:16b}).

(To be more presice the conditions $(\ref{eq:14})=
(\ref{eq:15})=0$ have other branches of possible solutions other than the
Abelian configuration. Other branches are (1) $u=0$ and $v=0$, (2) $a=fv$,
$D_{{z}}f=0$ and $a=hu$, $D_{\overline{z}}h=0$ where $f$ and $h$ are
real functions, and (3) the mixture
of (1) and (2). These branches are excluded in the following way. In the
first case (1) $ab=1$ and hence $a~(b)$ can not vanish on the Riemann surface.
Since $a~(b)$ is a charged scalar field, this forces the $U(1)$ bundle of the
gauge potential $A_{i}$ to have a vanishing degree. This is not possible
in the generic situation. In the second case (2) we obtain $D_{{z}}f=
D_{\overline{z}}f=0$ $(D_{\overline{z}}h=D_{{z}}h=0)$ since $f~(h)$ is real.
Hence $f~(h)$ is a
covariantly constant
scalar field. This is again possible only when the bundle is trivial. This
analysis is parallel to the case $SU(2)/U(1)$ discussed by Witten \cite{W3}.)

Thus the only contribution to the path integral
of (\ref{eq:10}) comes from the region around the singularity $uv\simeq 1$.
Fields $a,b,\alpha,\overline{\alpha},\beta,\overline{\beta}$ correspond to the
transverse directions to the sigularity and their path integrals cancel except
for an anomaly \cite{W3}. Thus our topological model in fact describes
only the region around the singularity in the $2d$ black hole geometry.
The black hole singularity is the chiral ring of the model (\ref{eq:10}).

 We notice that there is an interesting dualism
between the supersymmetric and topological
models (\ref{eq:8}, \ref{eq:10}). If in the supersymmetric theory
(\ref{eq:8}) we
solve the gauge fields in terms of the other fields and plug their values
back into the action, we obtain a supersymmetric non-linear $\sigma$-
model with the target space metric given by (\ref{eq:6}) (see Nakatsu in
\cite{Nak}). This procedure is legitimate except at the singularity.
Thus the
non-linear $\sigma$-model describes the target space geometry except for
its singularity. With the topological model describing only the region around
singularity
they nicely complement each other. While the map
of the non-linear $\sigma$-model explores the entire target space, the map
of the corresponding topological theory probes only its degenerate subspace.
(The possibility of a topological theory being constant maps onto
degenerate points in the target space has been discussed by Vafa \cite{Va}).

In a more general context we may speculate about the nature of TFT making use
of the idea of BRST fixed points. Suppose we have a TFT whose field space has
a non-trivial geometry and is covered by a number of gauge patches. In
each
gauge patch BRST transformation acts freely on the field space. Fixed points
of the transformation will possibly occur at the borders of the neighboring
patches. When the field space is interpreted as describing a space-time,
gauge patches will correspond to regions of space-time with different
causal properties and
BRST fixed points will occur at the space-time singularities.
At present we do not yet have an exactly soluble $2d$ field theory describing
realistic $4$ dim. black holes. When such models become available,
it would be extremely interesting to see
if in fact the singularities are described by some TFT as we have argued in
this article.

Putting aside the issue of the singularities let us now discuss the idea of
BRST fixed points and field configurations with a reduced holonomy in the
field space.
In the case of TFTs obtained from $N=2$
supersymmetric
filed theories by twisting, BRST fixed points are obtained from
the fixed points of supersymmetry transformations in the original $N=2$
theories.
Fixed points of supersymmetry are nothing but the supersymmteric vacua
and have been studied extensively in various model field theories.

In the case of $4$-dimensional $N=2$ supersymmetric Yang-Mills theory,
for instance, the supersymmetry fixed
point is given by the 't Hooft-Polyakov monopoles in the
Bogomolni-Prasad-Sommerfield limit
\cite{OW}. In fact the supersymmetry variation of a spinor field is given by
\beq
\delta \psi^{i} = F_{\mu\nu}\sigma^{\mu\nu}\epsilon^{i}
+ D_{\mu}A\gamma^{\mu}\epsilon_{ij}\epsilon^{j}
+ iD_{\mu}B\gamma^{\mu}\gamma_{5}\epsilon_{ij}\epsilon^{j}
+ \gamma_{5}[A,B]\epsilon^{i},~~i=1,2
\label{eq:19}
\eeq
and $\delta \psi^{i} = 0$ has a non-trivial solution when the gauge $A_{\mu}$
and scalar fields
$A,B$ form the configuration of magnetic monopoles and their contributions
cancel in (\ref{eq:19}).
In twisting the $4$-dimensional $N=2$ theory we redefine the Lorentz group
as $SU(2)_{L}\otimes
SU(2)_{R}'$
where $SU(2)_{R}'$ is the direct sum of $SU(2)_{R}$ and the internal symmetry
group
$SU(2)_{I}$. Under twisting,
fermions (doublet of $SU(2)_{I}$) are converted into ghost fields
$\psi_{L}^{i} \rightarrow \psi_{\mu}$,
$\psi_{R}^{i} \rightarrow \eta$, $\chi_{\alpha\beta}$ where
$\chi_{\alpha\beta}$ is self-dual
in its index $\alpha$, $\beta$.
Supersymmetry transformation law (\ref{eq:19}) is split into BRST
transformations
\begin{eqnarray}
&&\delta \chi_{\alpha \beta} = \epsilon F^{(+)}_{\alpha\beta},  \label{eq:20}
\\
&&\delta \psi_{\mu} = \epsilon D_{\mu}\phi,    \label{eq:21}  \\
&&\delta \eta = \epsilon [\phi, \phi^{*}]        \label{eq:22}
\end{eqnarray}
where $\phi=A+iB$ and $F^{(+)}_{\alpha\beta}$ is the self-dual part of the
field strength
$F_{\alpha\beta}$.
These are exactly the transformation laws of the topological Yang-Mills theory
\cite{W2}.
BRST fixed points are given by the anti-self-dual instantons (with $\phi=0$).

 There exists a close analogue of 't Hooft-Polyakov monopoles in $4$-dimnsional
gravity theory
which are the extreme Reissner-Nordstrom black holes (black holes with their
masses being equal to their charges). Extreme Reissner-Nordstrom solutions
occur
in $N=2$ extended supergravity theory (SUGRA) which is a supersymmetric
generalization of the coupled Einstein-Maxwell theory. In $N=2$ SUGRA
the local supersymmetry variation of the gravitino field is given by
\beq
\delta \psi^{i}_{\mu}=D_{\mu}(\omega)\epsilon^{i} + \frac{1}{2\sqrt{2}}
    \epsilon^{ij}(\hat{F}_{\mu\nu}\gamma^{\nu}+\frac{1}{2}e\hat{F}^{*}_{\mu
    \nu}\gamma^{\nu}\gamma_{5})\epsilon^{j}, ~~i,j=1,2
\label{eq:23}
\eeq
where $\omega$ is the spin connection, $F^{*}_{\mu\nu}$ is the
dual of $F_{\mu\nu}$ and $e$ is the determinant of the vierbein.
$\hat{F}_{\mu\nu}$ is the supercovariant Maxwell field strength
\beq
\hat{F}_{\mu\nu} = (\partial_{\mu}A_{\nu} - \frac{1}{2\sqrt{2}}\overline{\psi}
                     ^{i}_{\mu}\psi^{j}_{\nu}\epsilon^{ij}) - (\mu\rightarrow
                      \nu)).
\label{eq:24}
\eeq
It is known \cite{GH} that $\delta \psi^{i}_{\mu}=0$ has non-trivial
solutions
(Killing spinors) at the extreme Reissner-Nordstrom solution due to the
balance between the gravitational and electro-magnetic force. The
extreme Reissner-Nordstrom geometry has an unboken global $N=2$ supersymmetry
and is a close gravitational analogue of magnetic monopoles in
$N=2$ super Yang-Milles theory.
Actually there is a wider class of solutions to the supersymmetry fixed point
equation $\delta \psi^{i}_{\mu}=0$ and is called the Israel-Wilson metrics
\cite{IW}
in general relativity. Israel-Wilson metrics include extreme Reissner-Nordstrom
black holes in one end and contain self-dual gravitaional fields
(gravitational instantons) in the other end. It is quite curious to see if
we can again twist $N=2$ SUGRA and costruct a $4$-dimensional topological
gravity theory describing gravitaional instantons (coupled to self-dual
Maxwell fields). This problem is currently under study \cite{EK}.

We would like to thank Y.Nambu and the members of the theory group of the
Enrico Fermi Institute for their hospitality and interest in this work.
We also would like to thank G.Gibbons for discussions on the Israel-Wilson
metric and P.Horava on the parametrization of $SL(2;R)$.

\end{document}